# Rapid, parallel path planning by propagating wavefronts of spiking neural activity

Authors:

Filip Ponulak[1,3,*], John J. Hopfield[1,2]

[1.] Department of Molecular Biology, Princeton University. [2.] Institute for Advanced Study, Princeton. [3.] Institute of Control and Information Engineering, Poznan University of Technology, Poland

[*] Correspondence should be sent to: Filip.Ponulak@braincorporation.com.

## Summary:

Efficient path planning and navigation is critical for animals, robotics, logistics and transportation. We study a model in which spatial navigation problems can rapidly be solved in the brain by parallel mental exploration of alternative routes using propagating waves of neural activity. A wave of spiking activity propagates through a hippocampus-like network, altering the synaptic connectivity. The resulting vector field of synaptic change then guides a simulated animal to the appropriate selected target locations. We demonstrate that the navigation problem can be solved using realistic, local synaptic plasticity rules during a single passage of a wavefront. Our model can find optimal solutions for competing possible targets or learn and navigate in multiple environments. The model provides a hypothesis on the possible computational mechanisms for optimal path planning in the brain, at the same time it is useful for neuromorphic implementations, where the parallelism of information processing proposed here can fully be harnessed in hardware.

## Highlights:

- Hypothesis: brain rapidly solves path planning through parallel mental search
- Model: parallel mental search implemented by propagating waves of neural activity
- Planning solved using local plasticity rules during single passage of a wavefront
- The model can simultaneously embed multiple environments for trajectory planning

# Introduction

One of the central problems for neurobiology is to understand the computational effectiveness of the brains of higher animals. Brains rapidly carry out extraordinary feats of visual scene analysis or problem solving through thinking on 'wetware' that is tens of millions times slower than modern digital hardware. Part of the explanation is brute-force *anatomical* parallelism.

In this paper we develop a model of parallel *computational* processing in the context of path planning and spatial navigation. We propose that spatial navigation can be solved in the brain through simultaneous mental exploration of multiple possible routes. A typical mental exploration task for an animal might involve knowing an extensive terrain containing a few water sources, being motivated (being thirsty) to seek the nearest water source. Hopfield (2010) recently described a way that serial mental search for a useful route could be done by a moving clump of activity and synapse modification in a neural network. We show here that a best path can rapidly be found by *parallel* search in the same kind of network, but by a propagating wave of spiking activity. The resulting synapse modification pattern represents a vector field, which in turn can be used to generate motor commands leading the animal along the best route.

Can animals employ parallel mental exploration to solve novel problems? Indeed can humans do so? Published hippocampus electrophysiology experiments have not to date shown evidence of parallel exploration of possible actions. Mental activity trajectories related to actions soon to be taken have been found in rat hippocampus, but in highly overlearned situations (Ferbinteanu and Shapiro, 2003; Ainge et al., 2007; Johnson and Redish, 2007; Karlsson and Frank, 2009). Unfortunately there is no literature of electrophysiology experiments performed on animals solving a problem for which rudimentary thinking would be useful, namely while experiencing a situation for the first time. Classic rat behavioral experiments (Seward 1949) on novel problems can be explained on the basis of mental exploration, but other explanations might be possible.

One of the major roles of theory is to elucidate interesting consequences and possibilities inherent in our incomplete experimental knowledge of a system. The fact that hippocampus-like neural substrate can support parallel mental exploration, as explored here, is such a possibility. New experimental paradigms could easily test for parallel mental exploration in rats. These ideas also form the basis for novel neuromorphic circuits in engineering, which could also be used to implement effectively certain Artificial Intelligence algorithms such as those based on the idea of a wave-front propagation (Dorst & Trovato 1988, Dorst et al. 1991, LaValle 2006) by taking advantage of the through parallelism of the neuromorphic hardware systems (Boahen 2005, Misra & Saha 2010).

# Results

We consider like (Hopfield 2010) a network of excitatory 'place' cells for a very simple model animal. Through experience in an environment, each cell has learned synaptic connections from a sensory system (not specified here) that make it respond strongly only when the model animal is near a particular spatial location. These response place fields are our modeling equivalent of the response place fields observed in the rodent hippocampus (O'Keefe & Dostrovsky 1971). For display purposes, the activity of each place cell can be plotted at the spatial location of the center of the receptive field corresponding to that place cell. In such a display there is a localized activity clump surrounding the actual spatial position of the model animal. When the animal moves, this activity region follows the location of the animal. If an animal wanders throughout an environment over an extended time, the synaptic plasticity will result in excitatory synaptic connections being made only between cells that are almost simultaneously active (Hebb 1949). If the exploration process is not systematically directional and is extensive, connections will on average not have directionality. The CA3 region of the hippocampus has such intra-area excitatory connections with the requisite spike-timing-dependent plasticity, or STDP (Amaral & Lavenex 2006).

The fundamental neural network to be studied is thus a sheet of place cells, each having excitatory connections to the others with centers within its receptive field footprint, but not to distant neurons. Experimental support for the existence of such connections (direct or indirect) comes from the coordinated phase-change-like response of place cells, trained in two environments, experiencing a visual environment that mixes the two environments (Wills et al 2005).

The model neurons considered in our study are of the integrate-and-fire type with a short dead-time and spike-frequency adaptation (implementation details are provided in the 'Experimental Procedures' section at the end of the paper).

We investigate whether and how the described setup can implement parallel search for optimal pathways in the environment represented by the neural network. Because we rely on simulations of a system whose mathematics we cannot fully analyze, it is sensible to present a line of argument that develops insight about expected behaviors. Consider a simplified model comprising of a line of neurons, each reciprocally connected to its two nearest neighbors (cf. Figure 1). With specific parameter settings, a single spike can initiate an activity pattern that consists of a pair of spikes marching from the initiation site toward the ends of the line at constant speed, one in each direction (Aertsen et al. 1996). In a system with intrinsic neuronal adaptation, there is a dead time before another pair can be propagated in this same region.

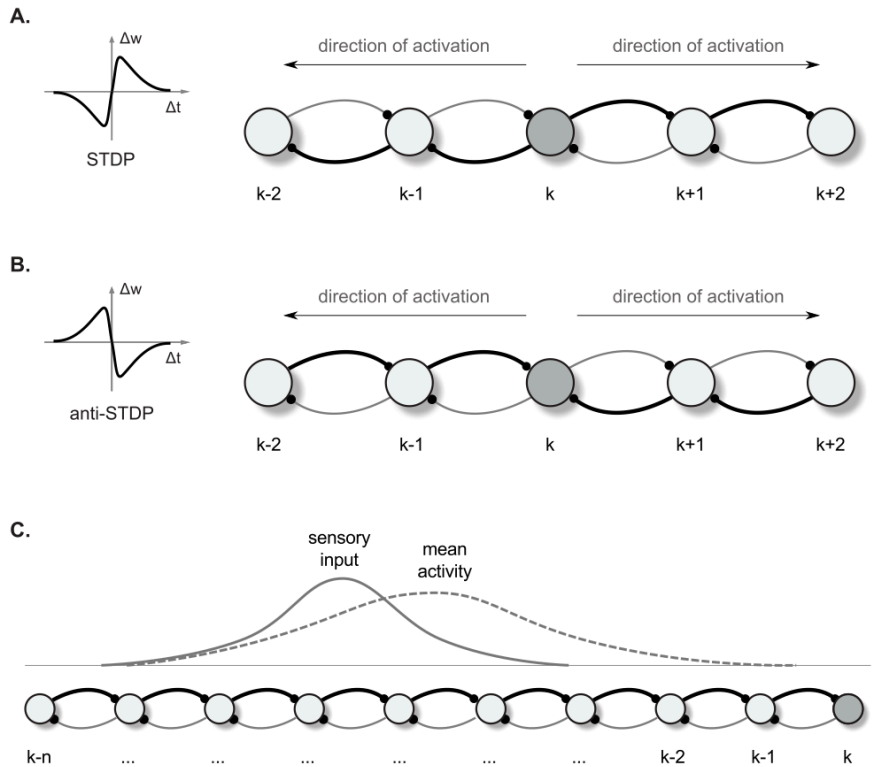

Figure 1: **Synaptic vector field formation.** (A,B) Illustration of the synaptic strength changes in a one-dimensional network altered by 'causal' STDP (A) and 'anti-causal' STDP (B) after a neural activity was propagated from the neuron k in the directions denoted by the arrows. The connections are shown as arcs with the direction of connection denoted by little dots representing synapses. Stronger connections are represented by the thicker lines. Left panels are the schematic illustrations of the synaptic weight changes $\Delta w$ as a function of the time lag $\Delta t$ between the post- and presynaptic spikes, for STDP (A) and anti-STDP (B). (C) Due to the asymmetry in the strength of connection from- and to- any particular neuron in the network, the mean neural activity observed in the network is shifted with respect to the input current distribution.

A similar phenomenon can be observed also in a two-dimensional sheet of neurons with recurrent local connections over a small but extended region. In an example presented in Figure 2.A and B, the synaptic connection strengths are chosen so that a few pre-synaptic cells must spike almost simultaneously to fire the post-synaptic cell. Seeded with a few approximately synchronized firings of nearby cluster of neurons, a propagating circular wavefront of activity is observed in which each neuron fires only once (Diesmann et al. 1999, Kumar et al. 2008). A second wavefront cannot be initiated in a region that the initial wavefront has traversed until the adaptation has decayed (cf. Figure 2.C,D). Note, that although in our model we consider a single-spike activity, the basic activity events propagated through the network may in principles also consist of short bursts of spikes, which is biologically more realistic in the context of the hippocampal cell activity.

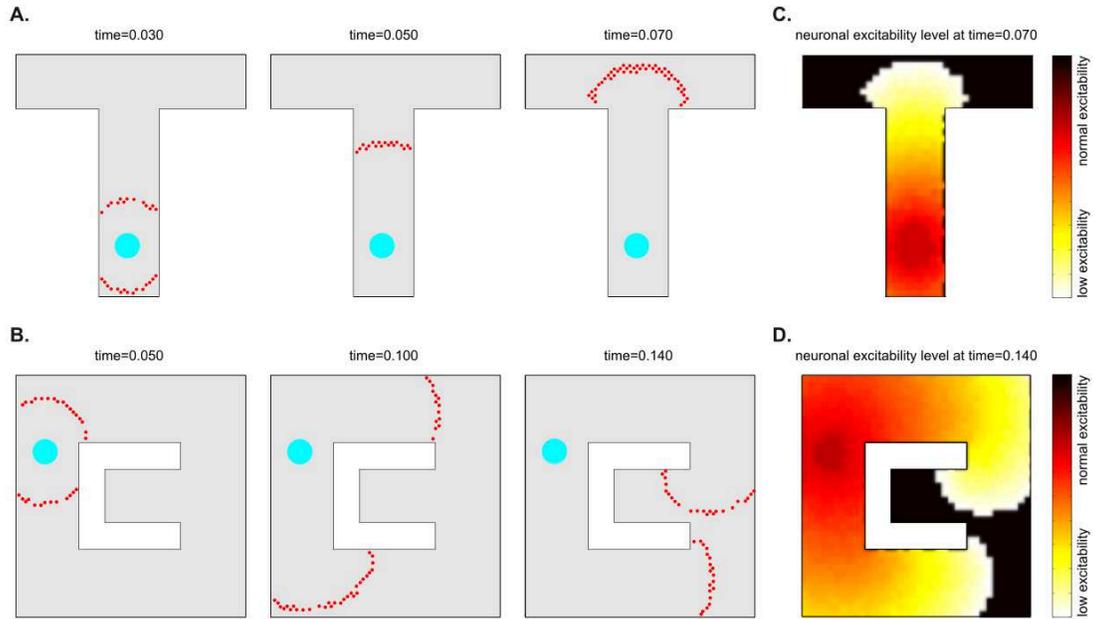

Figure 2: **Wavefront propagation and neuronal adaptation.** Illustration of a wavefront propagation in a network of synaptically connected place cells for two different environments (A) and (B). Cyan fields are the initiation points of the wavefronts. Red dots are the action potentials that occurred in a time window of 0.002 sec. centered at the times indicated. Plots C, D show color maps of the average level of a neural adaptation in the particular regions of the network after a single wavefront passage up to the states illustrated in the right-far plots in A,B, respectively. Brighter colors in these maps represent lower excitability of the neurons at the corresponding locations.

Propagating wavefronts can have profound effects on synaptic modifications through STDP. Consider again a one-dimensional network as illustrated in Figure 1. Any non-symmetric STDP rule will produce, in one dimension, synaptic change patterns that display whether the 'front' of activity that went by was going toward the left or toward the right. Normal or 'forward' STDP which enhances synapses at which the pre-synaptic spike comes before the post-synaptic spike will result in rightward-going synapses being stronger than leftward-going synapses if the wavefront passes moving to the right (Figure 1.A). 'Reverse' or 'anti-' STDP which enhances synapses at which the pre-synaptic spike comes after the post-synaptic spike (Bell et al. 1998, Kampa et al. 2007, Roberts & Leen 2010) will result in leftward-going synapses being stronger than rightward-going synapses if the wavefront passes moving to the right (Figure 1.B). The same basic idea intuitively extends to two dimensions, where STDP results in synaptic change that can be interpreted as a vector field (in the following we shall call it a synapse vector field or SVF), showing the orientation of the propagating wavefront that caused the synaptic change. In all our simulations we use reverse STDP induced by propagating spike wavefronts that creates an SVF pointing towards the center (initial point) of the waves. One could also use regular STDP that will result in an SVF oriented away from the center point. Our use of reverse STDP is motivated by certain conceptual and technical advantages of this approach over regular STDP, as it will be described later in the paper.

**Simple path planning problem**

Consider for definiteness the 'T' shape environment shown in Figure 2.A. We presume that by exploring the environment, each neuron has acquired a place field such that it is driven strongly only when the simulated animal is near the place field center and the drive to the cell falls off smoothly away from that location. For display purposes, in all figures the cells are arranged so that whatever property of the cell is being plotted, its *(x,y)* plot location is the location of its place field center. The receptive fields considered in our experiments are assumed to have Gaussian shapes and to cover 25-50 cells in their footprints in a simulation using a network with 2000 place cells. In such a setup, if an animal explores an environment, synapses with simple STDP will form strong connections between neurons with similar place fields, i.e. between neurons that are close together. To this point, the general approach is like that previously used in Hopfield (2010).

Imagine that the simulated animal, in exploring an environment, finds a target T, such as a source of water, to which it may later want to return. Let the dendrites of the place cells in the vicinity of T become connected to axons from an 'exciter' which, when activated, can briefly drive these place cells to fire. Such activation will result in an outgoing wave of single spike activity emanating from T as center as illustrated in Figure 2.A (where the cyan field represents the T location). This wave will spread until every neuron has fired an action potential. As noted before, the next wavefront is possible only after the neural adaptation fades away.

The propagating wave and the asymmetric synaptic plasticity implicitly define a vector field, which represents the local direction of the wavefront, i.e. the vector is normal to the wavefront and points in the direction of propagation. Sample SVFs that result from the anti-STDP rule are shown in Figs 3.A-D. Here the vector fields are illustrated using directed arrows originating from the preferred locations of each place cell in the network. The direction and the length of each arrow represent, respectively, the direction and the strength of the vector field in a given location (for more details we refer to the 'Experimental Procedure' section). If normal STDP had been used, the SVF would simply be reversed.

The synaptic vector field can be used for finding the shortest pathway to the location being the source of the propagated wave. Intuitively, since the first wave to arrive at your position comes via the fastest path, if you simply backtrack, always going backward along this vector field, you will reach the target by the shortest path.

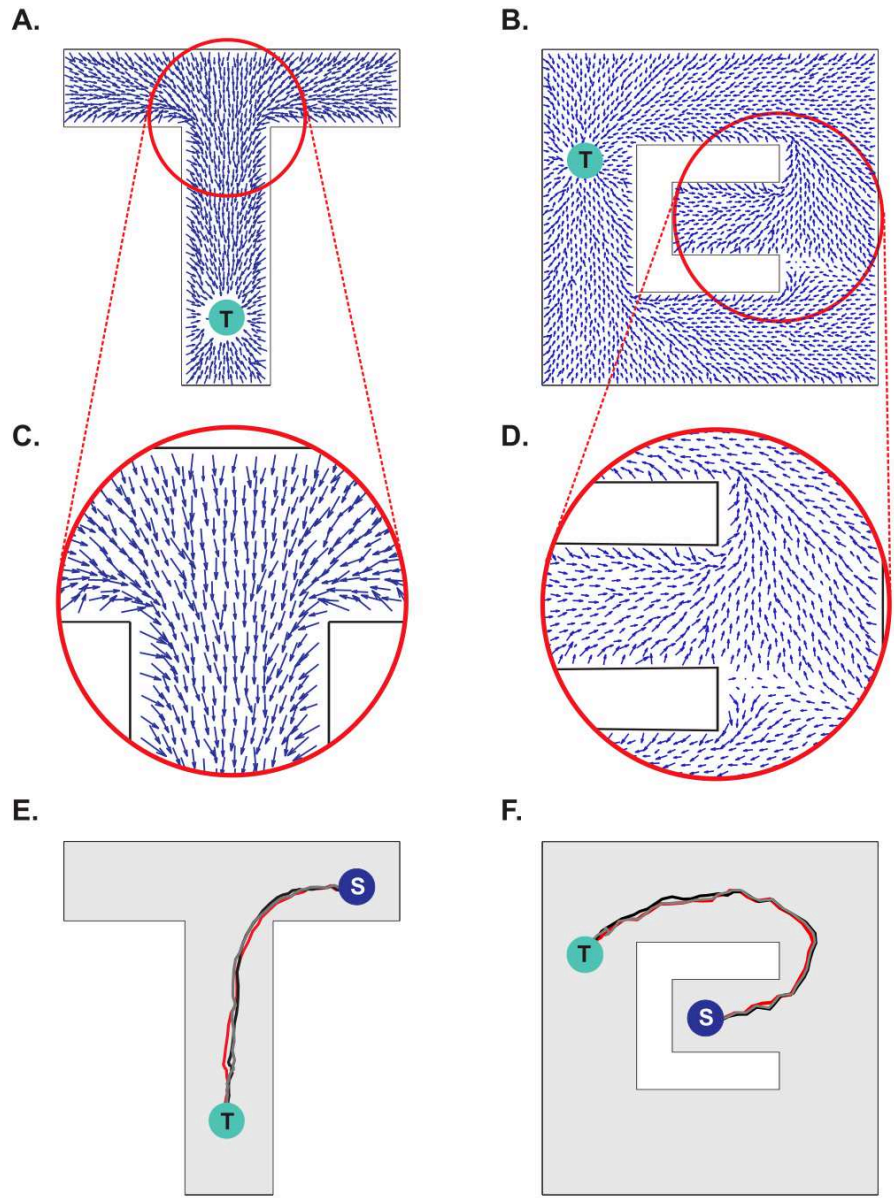

Figure 3: **Synaptic vector field and spatial navigation.** (A,B) Synaptic vector fields resulting from the wavefront initiated at point T and propagated as illustrated in Figures 2.A and B, respectively. (C,D) The insets show details of the vector fields around the bifurcations in the simulated mazes. (E,F) Typical movement trajectories observed in the considered models resulting from the vector fields from A and B, respectively. The trajectories begin in points 'S' and end in the target locations 'T'. For additional results see also Movie S1 in supplementary materials.

**Obtaining motor commands for following the synaptic vector field**

For illustrating an idea of how the synaptic vector field can be used for guiding an agent (a simulated animal or a robot) movement, we return to the one-dimensional case (Figure 1.C). In 1 dimension, if the propagating wavefront has passed by locations $k$ through $(k-n)$ while moving leftward, and the anti-STDP rule has been applied, rightward-directed synapses (e.g. $(k-1){\rightarrow}k$) are strengthened more than leftward ones ($k{\rightarrow}(k-1)$). Before this process, if the animal was located at a particular location in space, a bump of place cells would be active, symmetrically located around that location. In the presence of the asymmetric synapse modification, the bump of activity is biased and no longer centered on the actual physical location (cf. Figure 1.C). This bias can be converted into a motor command proportional to the bias and pointing towards the direction of a wavefront passage. Precisely the same 'move to improve the overlap between two blobs' problem occurred in earlier work (Hopfield 2010), where a fully neural model for converting the blobs disparity into the appropriate motor commands has been proposed. Since the task of generation of motor command is not the major focus of our paper, here we use a simplified approach. We assume that a receptive field corresponding to the present animal location is activated by applying tonic excitation to the corresponding place cells. Any place cell firing a spike causes a pulse of force moving an agent towards the preferred location of that cell. The asymmetry in the weight configuration around the receptive field results in a higher probability of firing of those adjacent place cells that are located along a direction of a vector field. As a consequence an agent moves to a spatial new location along the optimal pathway. The details of the algorithm are provided in the 'Experimental procedures' section at the end of the paper.

Sample movement trajectories resulting from applying the described procedure to a simulated animal are shown in Figure 3.E,F (see also Movies S1 and S2 in supplementary materials). These trajectories result from the SVFs illustrated in Figure 3.A,C and Figure 3.B,D, respectively. In particular, Figure 3.F illustrates the shortest path aspect of the available information; because the target is located above the midline, the wavefront arrives at the branch containing the animal at S from above before the wavefront from below (cf. Figure 2.B and Figure 3.B). Neural adaptation prevents the wavefront arriving from below from penetrating this region. Thus the SVF leads to a route from S to T going upward.

Notwithstanding the fact that the algorithm used here is not providing details on the possible neural implementation, it is important to emphasize that this control algorithm is purely spike-based. Each individual spike contributes to the agent behavior change. The average population activity pattern determines the mean movement trajectory along the vector field, whereas the particular spikes add some stochasticity to the behavior (reflected e.g. in a small trial-to-trial variability of the movement pathways observed in Figures 3.E and F). Such stochasticity has some advantageous in certain situations. For example, it may be useful for avoiding local minima, or for selecting one choice when several alternatives have equal probability.

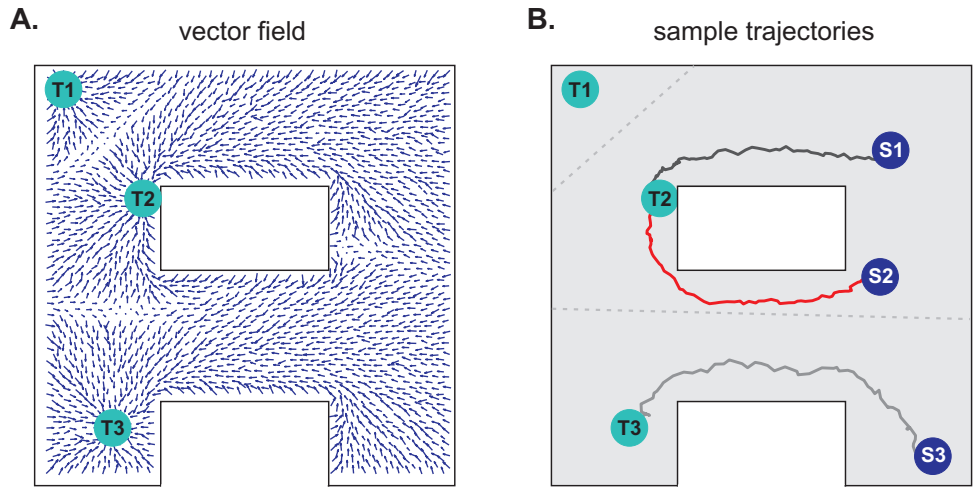

Figure 4: **Navigation in a system with multiple targets.** (A) Synaptic vector field created in the network with targets in locations T1, T2, T3. (B) Typical movement trajectories observed in the system for the initial agent locations as indicated by spots S1, S2, S3. The path selection and the path shapes are determined by the shape of the vector field and by the initial agent location's. The vector field has three basins of attraction corresponding to the particular targets - the bounds of the basins of attraction are indicated by the gray dotted lines. For additional results see also Movie S2 in supplementary materials.

**Navigation in an environment with multiple targets and values**

Several different relevant targets might be simultaneously available in an environment. For simplicity, the case when all targets have the same intrinsic value is first considered. Figure 4.A. shows the SVF that results when single spike propagating circular waves are simultaneously originated at three targets. Because the single-spike wavefront cannot propagate into a region that another wavefront has traversed, any subregion is therefore traversed by only a single wavefront, the one that arrives first, and is thus closest to its source. Within that subregion, the vector field is the same as it would have been if only the source responsible for the traversing wave had been present. The three subregions of the three possible targets of Figure 4.A are shown in Figure 4.B (compare to Movie S2 in the supplementary materials). Which target is nearest, and thus should be navigated to, depends on the current location of the agent. The same figure illustrates the paths followed for three possible initial agent locations. Note that the SVF is defined everywhere, independent of the location of the agent when the wavefront is generated.

If the targets all have equal value, the described procedure selects the target that can be reached by the shortest possible route. If there is a single intrinsic reward R for reaching a target, and a navigational cost of C per unit pathlength traversed, then the behavior generated maximizes $(R - C^*)$, where $C^*$ is the minimal path length from current location to target $T_k$ if $T_k$ alone is present, and where the maximum on k is taken over the real-space path lengths to all possible targets.

Now suppose that different possible targets $T_k$, k=1,... have different rewards $R_k$. The optimal behavior is now to choose the target on the basis of maximizing $(R_k - C^*)$. When all wavefronts propagate with the same velocity, it is useful to think in terms of times rather than lengths. $R_k$ then can be seen as an effective shortening of the time to navigate to a reward. Let $R_1$ be the greatest reward. Initiate a wavefront at $T_1$ at time t = 0. Initiate wavefronts at other target locations at times $(R_1 - R_k)/C^*v$, where v is the velocity of the wavefront as referred to real space. The introduction of these differential delays represents the value differentials between the various targets. These delays shift the boundaries of the regions such as those of Figure 4.B in a way that represents the differing values of the target. The optimal target is chosen. The path followed to that target is the same as would have been used if that target alone were present.

**Dealing with noise**

Noise can adversely affect the ability of the network to propagate a wavefront in the ideal fashion to set up the desired synaptic field. Figure 5 illustrates what can happen when noise is severe. Spurious single spikes are generated, and spikes can fail to occur. When spurious spikes cluster, they can serve as initiation sites for new circular waves centered at locations where there is no target. In addition, spurious and absent spikes cause irregular wavefront propagation or even wavefront extinction.

The major noise issues concern setting up the SVF. Once it is set up, the motor control system effectively averages over the vector field in a small region, and noise in following the SVF is not a major issue.

Having a large system is the first defense against noise and there is considerable latitude for exploiting the large number of cells available in real neurobiology. There are also cellular means to suppress the effect of noise, and we describe the effect of two of them in this section. Set the threshold for spike generation at some particular level, and consider the ability of *N* cells connected to this one to trigger it to spike when a passing wavefront goes by. Reliable wavefront propagation is enhanced by any biophysical effect that sharpens this threshold on *N*.

The first means of sharpening the threshold on *N* involves dendritic non-linear summation. In order to solve the problems associated with the threshold methods, a mechanism is required to determine whether a particular neuron in a network is excited by spikes coming from a small number of neurons being unusually effective because of noise, or by a larger number of neurons with typical effectiveness. A method of making this distinction can be implemented in a biologically realistic way by using supra-linear spatial summation, a phenomenon observed in biological neural circuits (Nettleton & Spain, 2000; Urakubo et al., 2004).

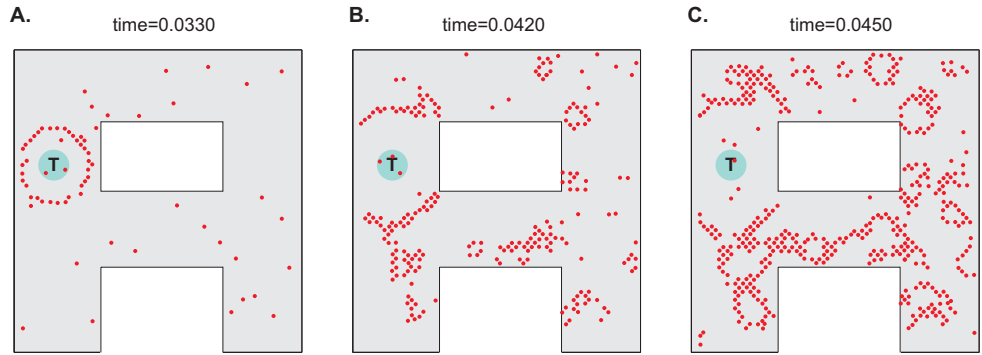

Figure 5: **Effects of noise on wavefront propagation.** (A) A single wavefront is initially started from the point T. Noise results in spurious single spikes or missing spikes. When spurious spikes cluster, they can serve as initiation sites for new circular waves centered at locations where there is no target. In addition, spurious and absent spikes cause irregular wavefront propagation or even wavefront extinction as illustrated in B and C. Network activity shown at times as indicated. The noise is modeled by injecting spike currents to randomly selected neurons at random time steps.

The algorithm for supra-linear summation used in our paper is presented in the Experimental Procedures section. Although, in this algorithm the appropriate setting of the neuron activation threshold is still important, it is no longer a critical factor for the problem at hand. With this approach more emphasis is put on how many presynaptic neurons are active simultaneously, rather then how strong the particular connections are. In this way the algorithm works better than the threshold algorithm for networks with greater heterogeneity of synaptic connection strengths.

While the supra-linear summation greatly reduces the problem of extra waves initiated by isolated spikes, it aids stable propagation of the existing waves of activity only in a limited way. The problem of stable propagation of neural activity becomes pertinent whenever the wavefront density changes. This is the case for example when the shape of the part of the environment recently explored by the wavefront changes (cf. Figure 6.A-D). As a consequence, as the wavefront travels through the network, the number of cells active near-simultaneously changes and so does the result of the integration function in the postsynaptic neurons. In this way the probability and the timing (the time lag between the stimulus and the response) of the postsynaptic firing varies for the various regions of the network. This effect, together with the effect of noise (from potentially multiple sources), results in a trial-to-trial variability of the profiles of the whole network activity observed as the wavefront travels through the network (cf. Figure 6, blue plot).

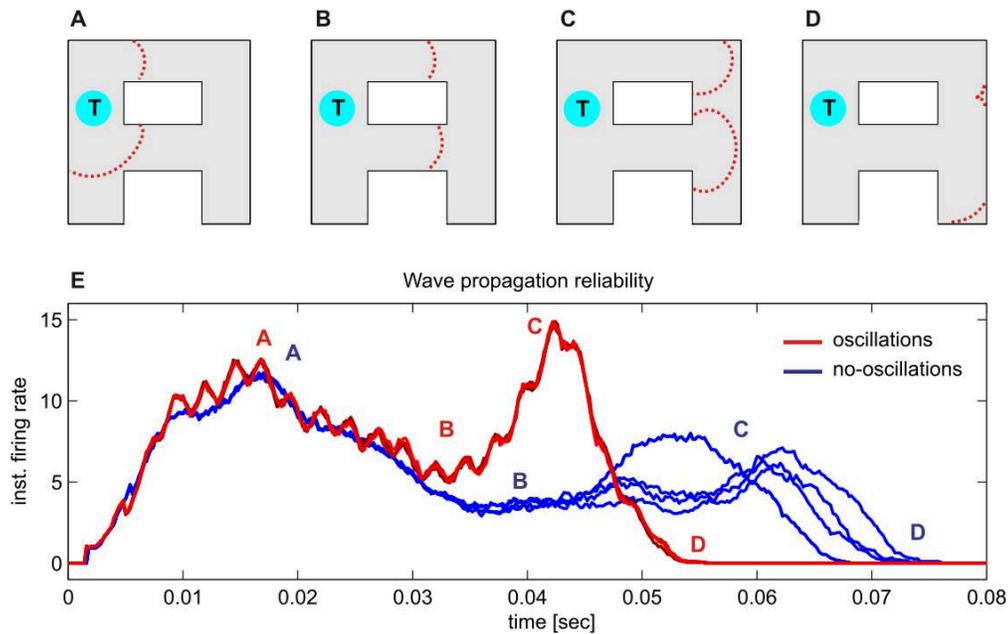

Figure 6: **Fast background oscillations increase wavefront propagation reliability.** Top panel: The wavefront is initiated in a certain environment at point T. Four screenshots of the different wavefront states are shown in A,B,C,D. Bottom plot: Reliability of the wavefront propagation can significantly be improved by driving the network with high frequency background oscillations (here with frequency of 400Hz). This reliability is evaluated by observing the variability of the population firing rate profiles during the wavefront propagations across different trials. In the experiment the wavefront is propagated several times starting from the location T. The same initial conditions are assumed for each trial. Four instances of the population firing rate profiles for the system with no-oscillation (in blue) and with background oscillations (in red) are shown. Population firing rates corresponding to the wavefront states shown in the top panel are indicated by the letters A,B,C,D, respectively. In case of the network with no-oscillations the firing rate profiles are significantly different in every trial. In case of the network with background oscillations the firing rate profiles are highly similar in every trial indicating high reliability of the wavefront propagation.

We found that the wave propagation process can significantly be stabilized and its trial-to-trial variability strongly reduced if the network is also driven by a background oscillation. Indeed, in the presence of the sub-threshold oscillatory drive, the variability of the temporal profiles of the network activity is chiefly suppressed. Figure 6.E (red plot) illustrates such suppression of noise-induced propagation irregularities by a high-frequency sub-threshold current injected into all the neurons. This reduction of the trial-to-trial variability of the wavefront propagation is manifested by the almost perfect overlap of the network activity profiles observed in 4 different trials. The effect of the subthreshold current is to effectively discretize time in units of (subthreshold frequency)$^{-1}$ and to synchronize all neurons firing within this short time window. The effect of small fluctuations in timing is suppressed, and of larger fluctuations exaggerated. Exploiting this effect requires a subthreshold period comparable to the cellular integration time. Interestingly, high frequency background oscillations are observed during the hippocampal sharp-wave ripples (Ylinen et al. 1995, Csicsvari et al. 1999), being the neural events linked with memory reactivation, exploration or planning (Wilson & McNaughton 1994, Jackson et al. 2006, O'Neill et al., 2006, Hopfield 2010). Thus our model suggests an explanation for the functional role of such high-frequency oscillations in the hippocampal areas as a means for stable oscillation of the propagated waves of neural activity, if the neuronal waves are indeed used for planning and navigation.

**Navigation in multiple environments**

When a rat is familiar with multiple environments, a particular hippocampal neuron can have place fields in more than one environment, with no apparent coordination between them (Bostock et al. 1991, Wilson & McNaughton 1993). We also therefore investigate whether the same network can learn and effectively perform navigation in multiple environments when each neuron has a place field in each environment. When the place cells in one-environment and place cells for a second environment are uncorrelated, the synaptic connections needed in both environments can be simultaneously present. If the number of neurons is sufficiently large, when the sensory signals come from one environment there is little crosstalk between the representations of both environments, and the presence of the second set of synapses simply inserts a modest level of noise. One can similarly anticipate that single spike wavefronts can be initiated and will propagate in any particular environment when multiple environments are known. The wavefronts will produce a vector field that can later be used to guide the animal in this particular environment. This is a significant extension, for without it each neuron needs to be specific to a single environment, which would be both inefficient and not in correspondence with biology.

Consider a network that is supposed to operate on two different environments as illustrated in Figure 7. Due to their shapes we call these environments 'A' and '∞'. While in the rat many place cells would be specific to one environment, such specificity reduces the crosstalk between the environments, and de-emphasizes the crosstalk effect we wish to evaluate. Here, however, we assume that each place cell represents the animal's locations in both environments.

A spike generated in any cell will produce excitatory postsynaptic potentials in all its neighbor cells in one environment and all its neighbor cells in another environment. As in the previous experiment, the model parameters are set such a single spike cannot cause action potentials in the postsynaptic neurons. As before, supra-linear summation helps to promote stable propagation of the existing wavefronts, and to prevent single, isolated spikes from producing new wavefronts. Also the high-frequency background oscillations are used to improve stable propagation of the neural wavefronts.

Consider a network activity caused by the simultaneous excitation of a certain set of the topologically nearby cells in the environment 'A'. When a plot is made with each cell located at the preferred location it represents in environment 'A', the dynamics of this neural activity will be seen as a wave propagated through the network (Figure 7.A,left).

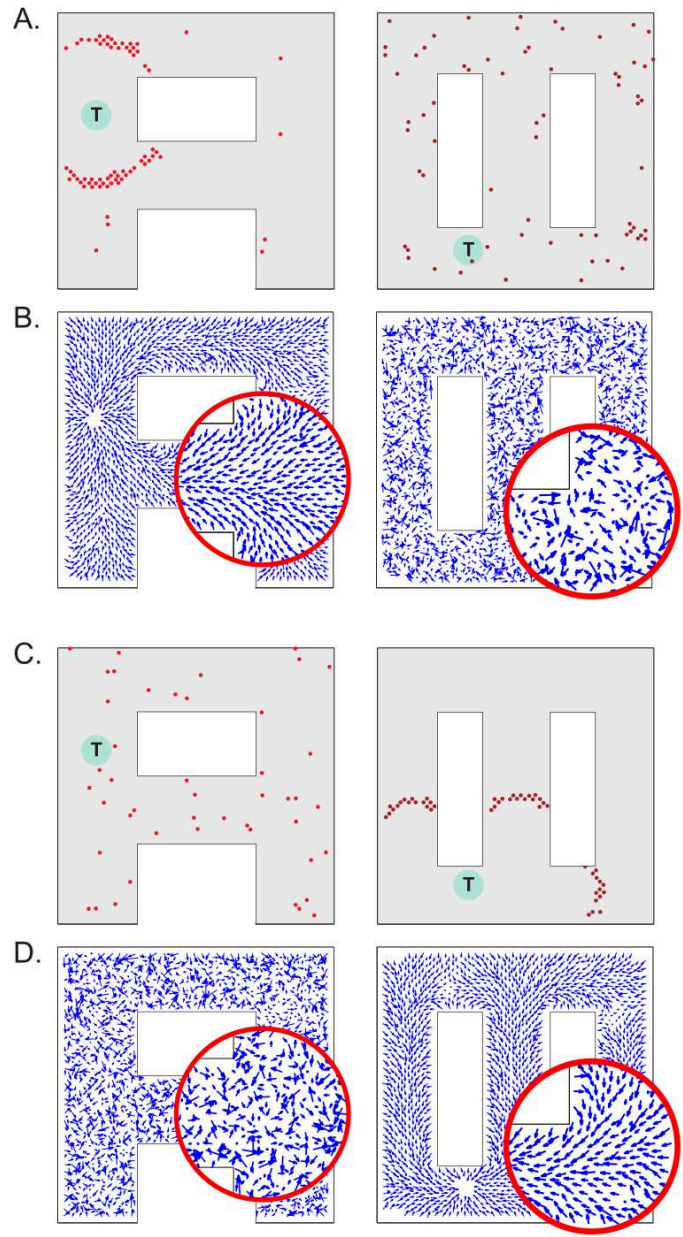

Figure 7: **Synaptic vector field formation in multiple environments.** (A,left) Wavefront propagation in environment 'A' short time after the activity wave initiation at the target T. (A,right) The same activity pattern as in (A,left), but displayed in the '∞' environment plotting representation. (B,left) Synaptic vector field resulting from the propagation of a wavefront illustrated in (A, left). Note a single attractor corresponding to the location T, that is the center of the wavefront. (B,right) The synapse vector field due to the same synapse changes as in (B,left), but calculated using the positions of the neurons in the '∞' environment. (C,D) The same plots as for (A,B) except that the wavefront has been initiated at target T in the '∞' environment. All results are qualitatively like those in A and B, except that the roles of the two environments are reversed. Synaptic vector fields in plots B and D are visualized using the same normalization factor (arrow scale) for both environments.

The same activity observed from the perspective of the '∞' environment (that is by reorganizing the network by putting place cells at the locations they represent in the '∞' environment) would appear as a random network activity (Figure 7.A,right). Since the spikes observed in the '∞' environment appear sparse, they are unlikely to initiate a wavefront in this representation. Similarly, at any particular moment while a wavefront in the 'A' environment is propagating, the synaptic connections representing the '∞' environment introduce drive to neurons that should not be driven at that moment. Occasionally such neurons can produce crosstalk-induced spurious spikes (cf. solitary spikes in the left panel in Figure 7.A, occurring far away from the wavefront).

Figure 7.B-D illustrates that at the level of two environments and around 2000 place cells, there is little effect of crosstalk on the ability to function in each environment as though the other did not exist. Figure 7.B (left) shows that the SVF induced by a wavefront initiated at T (cf. Figure 7.A,left) develop as expected, representing a flow back toward the target from all points in the 'A' environment. Figure 7.B (right) shows the SVF for the same synaptic changes, but calculated for the place cell locations in the '∞' environment. Here the vectors point in random directions because there is no spatial organization to the synapse change in this representation. The same kind of result is obtained when the wavefront is initiated in the '∞' environment as in Figure 7.C-D with the roles of the two environments reversing. In each case the vector field created by the single-spike wavefront successfully navigates an animal from a starting point in the given environment to the target as illustrated in Figure 8.

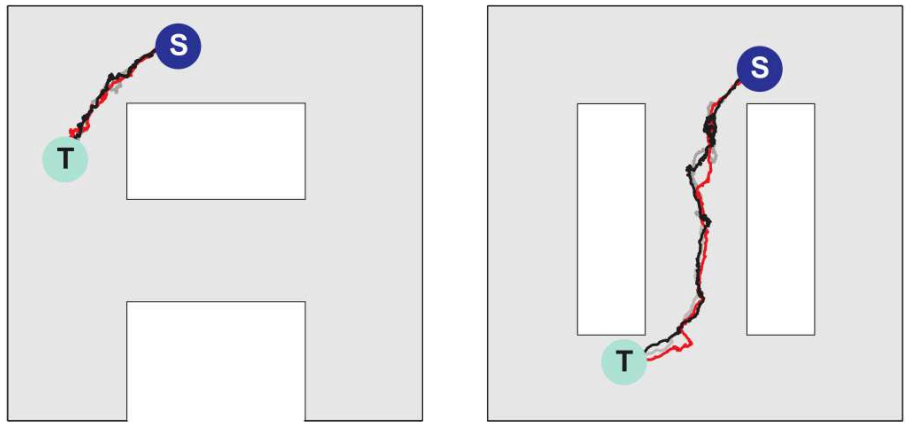

Figure 8: **Navigation in multiple environments.** Sample movement trajectories in the environment '$A$' (left panel) and '∞' (right panel) resulting from the synaptic vector fields shown in Fig.7.B(left) and Fig.7.D(right), respectively. Three different trials for each environment are illustrated. The trajectories start in the S locations and end in the T locations.

## Discussion

The problem of planning and executing a complex motion over a protracted time-period which will optimally take an autonomous agent from its present location and configuration to a desired target location and configuration is common to both animal behavior and robotics. In its simplest manifestation there is only a single target, a single known environment, and a short or fast path is preferred over a longer or slower one. The trajectory planning must accommodate the physical constraints posed by the environment. Additional complexities might include the simultaneous presence of multiple targets, possibly of different intrinsic values, terrain which affects the value of trajectories in a non-trivial fashion, and multiple environments.

The wave-propagation concept has first been introduced by Dorst and Trovato as an efficient parallel method for path planning (1988; Dorst et al. 1991) and since then has widely been used in robotics and computer science (LaValle 2006). The wave-front methods are essentially the same as exhaustive or heuristic versions of a classical A* search algorithm (Hart et al. 1968, Dijkstra 1959) of whose optimality is proven. Several neural models for spatial navigation using the concept of propagating waves have been proposed so far (for reviews see e.g. Lebedev et al. 2005, Qu et al. 2009). However, only a few models addressed a question on how the propagating neural activity can be transformed into an appropriate configuration of synaptic connectivity able to later guide an agent to a target location (Roth et al. 1997, Qu et al. 2009). To the best of our knowledge, our model is the first one to demonstrate that biologically plausible, temporally asymmetric synaptic plasticity rules can achieve this goal. Also, most of the previous models assumed multiple trials for learning a complete set of optimal paths for every new selected target location. In contrast, in our model, once an agent becomes familiar with an environment, a single passage of an activity wavefront through the network is sufficient to create a synaptic vector field guiding an animal from any possible location in the experienced environment to a target location.

The neurally-inspired network presented in our work has been shown to solve the planning problem in three steps.  First, in an exploratory phase it learns an environment by developing a set of 'place' cells whose locations reflect all possible trajectory boundaries due to kinematic constraints or constraints in the behavior arena.  It develops in this exploration process interconnections between all pairs of places that can be visited in temporal contiguity, and thus can be possible candidates for a section of a trajectory.

Second, given the expected set of synaptic connections, the excitation of a target location (or locations) initiates a wavefront of single spike- or single burst activity that propagates outward from the initiation site(s). The wave propagation process is terminated when a wavefront reaches the present location of the agent. The passage of such a wavefront produces synapse modification pattern that can be described as a vector field.  The desired trajectory is simply the path along the synaptic vector field from the present location to the (or a) target.  Since the synaptic vector field lines are produced by an expanding circular wavefront, they converge when followed backward toward a source, and thus provide stable guidance for going to a target location.

For use in robotics, a mathematical algorithm, like the one described in our paper, can be used to compute the optimal path from the SVF, and that optimal path can then be handed off to a guidance system for robot motor control. By contrast, for animals a means must be found to use the synapses themselves to physically traverse (or perhaps as a useful preliminary to mentally rehearse) that optimal path.  For the SVF is only our mathematical construction, and is not explicitly represented anywhere in an animal nervous system. Thus if these ideas are to have relevance to animal behavior, a third element is essential, a neural circuit or neurally-based method to obtain appropriate motor commands to robustly follow the optimum path, based only on the synapses themselves. The basic problem is to generate motor forces which will bring into better alignment two 'bumps' of neural activity, one coming indirectly from the sensory system representing the actual location of the agent, and the other clump of neural activity having a location biased by the modified synapses. This problem can be solved by a biological neural network, for the problem is isomorphic to the problem of moving the two eyes so that the image of one bright spot is centered on both fovea (Ohzawa 1990, 1997).  A relatively inefficient but fully neural solution to this two-bump problem was given in (Hopfield 2010).

The full extent of the parallelism available in our concept is perhaps best illustrated in Figure 4.  The system simultaneously selects the closest target and the best route to that target from a single propagation of the exploration wave. Conventionally, a best path would be found for each target sequentially, using a serial algorithm to rate possible paths, and a choice of target then made between these optimal single-target paths.  The conceptualization of the parallel search method and the demonstration by simulation that best trajectories can be followed in neuromorphic simulation are the major accomplishments of this paper.

We have shown through simulations that even a disorganized and non-topographic structure like the hippocampus should produce the kind of wavefront propagation (visible only in a carefully chosen presentation) necessary to rapidly solve novel route-choice problems in multiple environments.  We postulate that a short-term, small, STDP-based synapse change results from this collective wavefront propagation. A robust collective property of the underlying neural network then implements behavior, which follows the optimal trajectory, eliminating the need for mental or physical sequential search.

Generation of directed connections requires asymmetric STDP rules. Such asymmetry in the STDP learning windows has been found in the synaptic connections between hippocampal cells, first in cultured cells (Bi and Poo 1998) and more recently also in slice preparations (Aihara et al. 2007, Campanac and Debanne 2008).

'Anti-' or 'reverse-' STDP, in which a pairing of a pre-synaptic spike that precedes a post-synaptic spike *decreases* the strength of a synapse (Bell et al. 1997, Kampa et al. 2007), was used in our model to produce the synaptic vector field.  A system that has only normal or 'pro' STDP will produce a synaptic vector field that differs from the 'anti' field in having all vectors reversed.  Aside from minor technical differences related to adaptation, this field could serve to guide behavior merely by reversing the sign of all motor commands.  There are, however, two important distinctions.  If parameters are set in the fashion of (Hopfield 2010) so that a clump of activity, once initiated by sensory input, is stable when sensory input is removed, that clump of activity will move, following the vector field. Thus when the 'anti' sign is used, the agent can rehearse mentally the chosen trajectory from its present location to the chosen goal. It could even, with slight elaboration, communicate a sequential list of way points. Such a natural behavior of mental rehearsal in sequential order from the starting point is not available with 'pro' STDP, for the clump of activity in this case moves away from the target. Initiating a clump of activity at the target location does not create an equivalent in reverse order because the vector field diverges from that point. Another advantage of using anti-STDP over STDP is apparent for navigation in the presence of neural noise or external perturbation (physical forces pushing the agent away from the original path). When using anti-STDP, flow field lines converge when looking toward the source of the expanding circular wavefront that generated the field. When following in this direction, nearby vector field lines all converge toward the same destination, so noise is attenuated by the following process and has little effect. When following away from a source, as would be the case for normal STDP, vector field lines diverge, the effect of a noise error is amplified, and effects of noise accumulate.

Regarding the neural adaptation, there are two important temporal aspects of that process. The adaptation must be prompt - this is to ensure that the wavefront remains a single-activity event (a spike or a burst of spikes) per neuron phenomenon. The adaptation must last long enough such that the new wavefront cannot propagate across the region recently visited by another wavefront (hence the two wavefronts meeting at a certain point inhibit each other at that point). Since the wavefronts cover the entire region within 100-200 ms, this merely necessitates that the timescale for the recovery from adaptation be longer than this, as is the case in typical neuronal spike-frequency adaptation (Storm 1990, Sah & Davies 2000).

Accordingly synaptic changes (resulting in the directed SVF) are needed only for the duration of the behavior, after which some entirely different behavior involving different possible targets and requiring a different set of synapse changes (and SVF) may be needed to guide behavior. At any time, only one vector field is needed, since the animal has at any time a single behavioral goal. Interference from previous vector fields directing other motions can be avoided if the synapse changes due to the passage of a wavefront are short-lived. Such forgetting is particularly useful in this regard since if the present vector field is needed for a protracted time, it can be rapidly refreshed by initiating another wavefront from the same targets. However, we know of no studies testing the presence of short term (and perhaps rapidly decaying) rapid STDP. 'Spine twitch' (Crick 1982) is a strong candidate for producing such rapid but transient short-term synapse modulation.

**Outlook**

Our model can be usefully expanded in many ways. Different costs can be associated with the particular pathways or spatial locations through the uneven distributions of place cells and/or uneven distributions of strength of synaptic connections. These will affect the speed and the shape of the particular wavefronts, and consequently will determine the boundaries of the basins of attraction and best path within each basin. The plasticity process itself could be controlled, in space or in time, by neuromodulators (Seol et al., 2007), shaping the vector field. The model suggests the importance of searching for short term STDP based on very few spikes, and the importance of experimental paradigms in which mental exploration would enable an animal to solve a novel problem that would otherwise require trial-and-error exploration.

From the application point of view our neural algorithm can be extended to the path planning problems in systems with more than two dimensions or in tasks with extra constraints, such as e.g. non-holonomic navigation, arm movement planning. Our model, as a particular implementation of the wavefront expansion algorithm, can also be used for solving variety of optimality problems from other domains then motor control (Dorst et al. 1991, LaValle 2006).

## Experimental Procedure

The place cell models considered in the paper have been simulated using adapting leaky integrate and fire neurons. The dynamics of the neuron models between spikes are defined by the following formula:

$$\tau_m \frac{du_m(t)}{dt} = -(u_m(t) - u_r) + R_m (i_{sens}(t) + i_{syn}(t) + i_{ns}(t) - i_{inh}(t) - i_{Ca}(t)), \quad (1)$$

$$\tau_{Ca} \frac{di_{Ca}(t)}{dt} = -i_{Ca}(t), \quad (2)$$

where $u_m(t)$ is the membrane potential, $\tau_m = C_m R_m$ is the membrane time constant, $C_m = 1nF$ and $R_m = 20M\Omega$ are the membrane conductance and resistance, respectively, $u_r = 0mV$ is the membrane potential at rest, $i_{sens}(t)$ is the sensory input, $i_{syn}(t)$ is a sum of the currents supplied by the particular excitatory synapses entering the given neuron, $i_{ns}(t)$ is the non-specific background current modeled as a gaussian process with zero mean and variance 5nA, $i_{inh}(t)$ is the global inhibitory current, $i_{Ca}(t)$ represents a neuron-specific inhibitory current that could be caused by calcium-activated potassium channels in real neurons.

The neuron produces an instantaneous action potential when $u_m(t)$ reaches a threshold of 10mV , and then $u_m(t)$ is reset to 0 and held at that value for 2 ms to produce an absolute refractory period. Each action potential produced by the neuron allows for a momentary burst of calcium ($Ca^{2+}$) ions to flow into the cell (through high-potential Ca2+ channels) and increments $i_{Ca}(t)$ upward. Calcium ions also leak out, with a characteristic time $\tau_{Ca}$ usually set at 1-5sec. Because $i_{Ca}(t)$ and the internal $Ca^{2+}$ ion concentration of the neuron are proportional, the adaptive effect can be written in terms of the variables $i_{Ca}(t)$, and the cellular internal $Ca^{2+}$ concentration is needed only to understand a possible mechanism of spike-frequency adaptation. The timescale of adaptation is set by the size of increment to $i_{Ca}(t)$ that occurs when a neuron spikes.

For the calculation of the total synaptic currents $i_{syn}(t)$ injected into the particular neurons we use a supra-linear spatial summation model (Nettleton & Spain, 2000; Urakubo et al., 2004). The model favors a near simultaneous activation of a neuron from multiple presynaptic neurons over the activation from a single neuron. This approach is supposed to decrease the probability of initiating random wavefronts arising from isolated spikes in the noisy network. The model for supralinear summation used in our simulations is described by the following equation:

$$i_{syn}(t) = a_{syn} \tanh\left(b_{syn} \sum_j H(i_j(t))\right) \sum_j w_j(t)\, i_j(t), \tag{3}$$

where $i_j(t)$ is the synaptic current of the $j$-th input; $w_j(t)$ is the synaptic strength of the $j$-th input; $H(x)$ is the step function ($H(x)=1$ for $x>0$ and $H(x)=0$ for $x\leq 0$); $a_{syn}$ and $b_{syn}$ are the positive constants. The particular synaptic currents $i_j(t)$ rise instantaneously and decay exponentially with a 25ms time constant.

Sensory currents $i_{sens}(t)$ for each place cell are modeled as having an isotropic Gaussian form around the center of the receptive field for that cell, with the same width and strength for each neuron. When modeling multiple environments, each cell has a receptive field in each environment, assigned randomly.

It is assumed that the modeled network contains a set of inhibitory interneurons whose function is to limit the total activity of the network. Because the inhibitory feedback is assumed to be global, and because this essential function is computationally trivial, its effect is modeled in a continuous fashion and using global variables rather than by using spiking interneurons. Hence the dynamics of inhibitory population are given by the following equations:

$$\tau_e \frac{di_e(t)}{dt} = -i_e(t) + a_e \sum_j \sum_f \delta(t - t_j^f). \tag{4}$$

$$\begin{cases} A_{inh}(t) \propto (i_e(t) - I_{e0}) & \text{if } i_e(t) > I_{e0}, \\ A_{inh}(t) = 0, & \text{otherwise.} \end{cases} \tag{5}$$

The variable $i_e(t)$ represents the input current to the inhibitory population from all excitatory cells in the network, whereas $A_{inh}(t)$ reflects the activity of the inhibitory population. According to (4) the current $i_e(t)$ decays with a time constant $\tau_e$ and is incremented by $a_e$ by each individual spike fired at time $t_j^f$ (with $f$-being the label of the spike) by any excitatory neuron $j$ in the network. The parameters $\tau_e$ and $a_e$ are positive and constant; a Dirac function $\delta(.)$ is defined as: $\delta(t)=0$ for $t\neq 0$ and $\int \delta(t)dt=1$. According to (5) the population activity $A_{inh}(t)$ is proportional to the current $i_e(t)$ with a firing threshold $I_{e0}$. Given the activity $A_{inh}(t)$, the global inhibitory feedback $i_{inh}(t)$ to every excitatory neuron in the network is assumed:

$$i_{inh}(t) = a_{inh} A_{inh}(t), \tag{6}$$

where $a_{inh}$ is a binary gating variable. The gating variable $a_{inh}$ is set to 1, and accordingly the inhibition is active, during the network exploration or during the navigation task; whereas $a_{inh}=0$ and the inhibition is deactivated during the wavefront propagation.

A fully connected network with excitatory connections has been assumed in all simulations, with all network connections being initially silent. A typical size of the simulated networks varied from 2000 to 4000 place cells in the particular experiments. The simulations were carried out using an Euler integration of the differential equations and a 0.2-ms time step.

**Synaptic plasticity**

Synaptic connections have been altered according to the STDP model described by the following equation (cf. Kempter et al. (1999)):

$$\frac{dw_{ji}(t)}{dt} = a + d \left[ S_i(t) \int_0^\infty a_{ij}(s) S_j(t-s) \, ds + S_j(t) \int_0^\infty a_{ji}(s) S_i(t-s) \, ds \right], \tag{7}$$

where $w_{ji}(t)$ is the synaptic coupling from neuron $i$ to neuron $j$, $a<0$ is the activity-independent weight decay, $S_i(t)$ and $S_j(t)$ are the pre- and postsynaptic spike trains, respectively. A spike train is defined as: $S(t)=\Sigma_f \delta(t^f -t)$, where $t^f$ is the $f$-th firing time. The terms $a_{ij}(s)$ and $a_{ji}(s)$ are the integral kernels, with $s$ being the delay between the pre- and post-synaptic firing times ($s=t^f_i-t^f_j$). The kernels $a_{ij}(s)$ and $a_{ji}(s)$ determine the shape of the STDP learning window. In our model we use exponential functions given by (8) to describe the STDP curve, however, other shapes are also possible.

$$\begin{cases} a_{ji}(-s) = +A_{ji} \cdot \exp(s/\tau_{ji}) & \text{if } s \leq 0, \\ a_{ij}(s) = -A_{ij} \cdot \exp(-s/\tau_{ij}) & \text{if } s > 0, \end{cases} \tag{8}$$

Here, $A_{ji}$, $A_{ij}$ are the amplitudes and $\tau_{ji}$, $\tau_{ij}$ are the time constants of the learning window. In our model we assume that $A_{ji}>A_{ij}>0$ and $\tau_{ji}=\tau_{ij}>0$. The parameter $d$ in (7) controls the polarity of the STDP process and can be linked to the concentration of specific neuromodulators known to be able to change the polarity of the synaptic plasticity in biological synapses (Seol et al., 2007). For simplicity, in our model $d=\{-1,0,1\}$. We assume that during the environment exploration phase $d=1$, and consequently the synaptic connections undergo STDP with a positive net effect (because $A_{ji}>A_{ij}$). During the wavefront propagaton phase: $d=-1$ and accordingly the synaptic connections are altered by the reversed STDP rule. No synaptic plasticity is assumed during the movement execution phase ($d=0$).

**Synaptic vector field illustration**

In Figures 4,5,8 we present sample synaptic vector fields created by the propagating activity wavefronts. These vector fields are illustrated using directed arrows originating from the preferred locations of each place cell in the network. The direction and the length of each arrow represent, respectively, the direction and the strength of the vector field in a given location. Here we describe an algorithm used to illustrate the vector field.

For each neuron $n_i$ in the network consider a set $N_{ji}$ of all neurons $n_j$ on which $n_i$ makes direct synaptic projections. Now for the neuron $n_i$ we define a vector $r_i(t)$:

$$\mathbf{r}_i(t) = \sum_j w_{ji}(t)(\mathbf{x}_j - \mathbf{x}_i) / \sum_j w_{ji}(t), \qquad (9)$$

We assume that the vector $r_i(t)$ begins in the preferred location $x_i$ of place cell $n_i$ and ends in a center of gravity of the preferred locations $x_j$ of the neighboring place cells $n_j \in N_{ji}$, weighted by the corresponding connection strengths $w_{ji}(t)$.

**Exploration algorithm**

An exploration procedure was used to establish a set of synaptic connections appropriate to the topology of a particular environment, based on earlier work (Hopfield 2010). The trajectory followed was a noisy straight line with constant speed, with a directional persistence length of the same scale as the largest dimension of an environment. The trajectory made a specular bounce when it encountered a wall. During this exploration the place cells had sensory inputs according to their spatial receptive fields. Place field centers were assigned on a regular grid, with Gaussian noise around those locations. Pre-post synaptic spike pairs were accumulated for each intra-place cell synapse during the exploration. The potential for synapse change was evaluated over these spike pairs with a weighting function $dw_{ji}(t)/dt = exp(-|t_i-t_j|/\tau_e)$ and used to select which synapses should be established. In the equation, $w_{ji}(t)$ is the strength of the synaptic equation from a presynaptic neuron $i$ to a postsynaptic neuron $j$; $t_i$ and $t_j$ are the firing times of the pre- and postsynaptic neuron, respectively; $\tau_e$ is the learning time constant. When the exploration is finished, each place cell $j$ was given incoming synapses of the same size to the set of $m$ neurons with the largest values of weights $w_{ji}$.

This procedure is insensitive to the details. Since any trajectory could be traversed in either direction, it will yield virtually the same set of synapses over a large range of parameters and variations in the form of $S$, as long as there is a net positive area under the curve $S$, and the exploration is extensive. The resulting connection matrix is similar to that which would be achieved by connecting each place cells to its $m$ nearest neighbors.

**Navigation algorithm**

Once a vector field is created, a simple motor control algorithm is applied for the animal navigation. The algorithm is performed in the following steps:

1. A receptive field corresponding to the present animal location is activated by applying tonic excitation to the corresponding place cells

2. A weak global, activity-dependent inhibition (cf. Eqs.4-6) is applied to suppress random spikes resulting from the background noise or from crosstalk between different environment representations.

3. Every spike observed in the network is supposed to act as an instantaneous attractor causing a pulse of force moving the animal towards the preferred location of the active place cell:

$$\mathbf{F}(t) = a_F \sum_j \sum_f \delta(t_j^f - t) \left( \mathbf{x}_j(t) - \mathbf{x}_a(t) \right), \tag{10}$$

$$\mathbf{H}(\mathbf{x}_a)\ddot{\mathbf{x}}_a + \mathbf{c}(\mathbf{x}_a, \dot{\mathbf{x}}_a, \mathbf{F}_{ext}) - \mathbf{F} = 0. \tag{11}$$

Equation 10 defines the force vector $F(t)$ caused by spikes generated by place cells active at time $t$. Equation 11 describes the dynamics of the animals movement in the physical world. Here $x_a(t)$, $x'_a(t)$ and $x''_a(t)$ are, respectively, the location, velocity and acceleration of the animal's center of mass (for clarity we omitted the symbol t in Eq.11); $x_j(t)$ - is the preferred location of the place cell $n_j$ ; as before, $t_j^f$ is the firing time of the $f$-th spike in neuron $n_j$; $\delta(.)$ is the Dirac function; $a_F$ is the constant gain, $F_{ext}$ denotes all possible external forces acting on the animal, $H$ is the inertia matrix and $c$ is a bias force (Craig 2004).

## Acknowledgements:

Most of this work was done by the authors during a stay of F.P. at Princeton University sponsored by Dr. Carlos Brody, whose generosity we gratefully acknowledge. The authors would also like to thank Dr. Piotr Skrzypczynski for useful discussions on vector-field based algorithms for robot navigation.

## Author's present address:

Filip Ponulak: Brain Corporation, San Diego, CA 92121, USA